# Closed Form Approximations For The Three Body Problem

A. B. Mehmood, U. A. Shah and G. Shabbir

Faculty of Engineering Sciences,

GIK Institute of Engineering Sciences and Technology

Topi, Swabi, NWFP, Pakistan

Email: shabbir@giki.edu.pk

**Abstract**

In this paper, an approach is developed to solve the three body problem involving masses which posses spherical symmetry. The problem dates back to the times of Poincare, and is undoubtedly one of the oldest of unsolved problems of classical mechanics. The Poincare's Dictum comprehensively proves that the problem is truly insolvable as a result of the nature of the instabilities involved. We therefore refute the idea of finding exact solutions. Instead, we develop closed form analytical approximations in place of exact solutions. We will solve the problem for the case when all the masses involved have spherically symmetric mass distributions. The method of solution would include the use of a single mass to replicate the effect of two individual masses on each body. The derivation of solutions will involve the use of the Lambert's wave function and the solution will comprise of the position vectors expressed as explicit time functions.

## Introduction

Having outlined the nature of our work in the abstract, we go on to provide a more formal definition of the problem that we choose to solve. The problem, by definition is to solve for the position vectors as time functions, of three gravitating masses when they execute free motion under each other's gravitational influence. The masses form an isolated system in free space, and given the initial position and velocity of each mass, their subsequent free motions are to be examined as accurately as possible. We now state and discuss some of the simplifying assumptions that will be used in solution of the problem. It will be noted that these assumptions will provide us with the luxury of deriving the solutions comfortably on a relative scale, and yet the accuracy of our solutions will be reasonable.

Firstly, we will assume that the gravitating bodies involved posses mass distributions that are spherically symmetric. The reason behind this idea is the fact that the three body problem finds most of its applications in celestial mechanics [1-4], involving planets and heavenly bodies as the gravitating masses. Moreover, these planets and heavenly bodies are known to posses mass distributions that are almost always spherically symmetric. As a result it would be impractical, (and perhaps not possible) to solve the problem taking into account any arbitrary mass distributions. Moreover, this assumption would also allow us to treat the gravitating bodies having finite volume, as point masses. This would of course be another simplifying step since the effect of each body would be quite accurately replicated by a point mass placed at the centre of mass of the body. Having incorporated this assumption into our problem, we would no longer show any concern for rotational motion of the bodies about their respective axes of rotational.

As a second assumption, we will consider solving the problem only for the case when the masses involved have position vectors that are not arbitrarily large and angular velocities that are small (i.e. specifically less than 1 radians per second). It should be noted here that the term 'angular velocities' refers to the rotation velocities of the position vectors and not those of the masses themselves. We believe this assumption to be quite reasonable since angular velocities of the position vectors encountered in the cosmos are normally much less than 1 radians per second. Moreover we can always scale the position vectors so that they never become arbitrarily large. Therefore this assumption, is also in close conformance with reality, and should not affect the accuracy of our results in a manner that is non negligible. The fact as to how this assumption will truly negligibly effect our calculations will be demonstrated more clearly when we solve the problem other approach can be found in [5-6]. Having provided a sufficient discussion on our assumptions, we now go on to present a diagrammatic representation of our system of three bodies. Figure 1 serves as an adequate aid for this task.

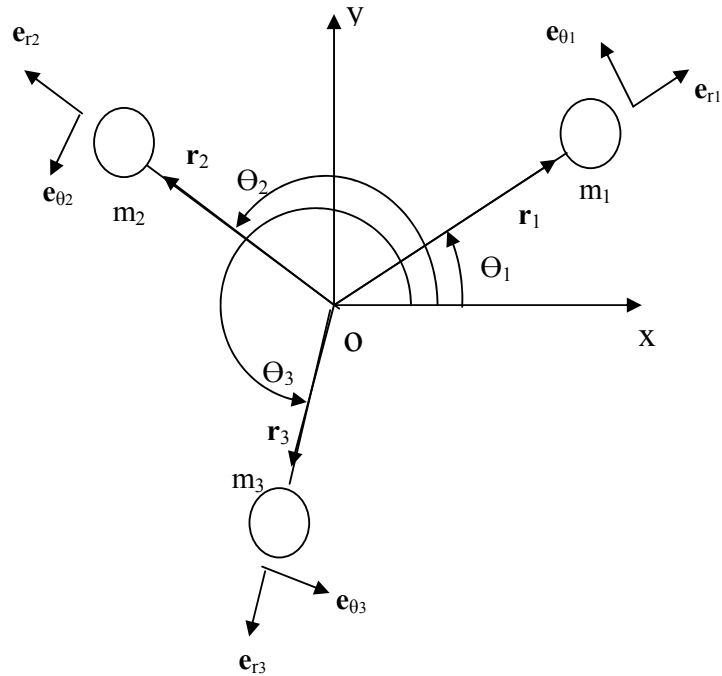

**Figure 1**

Here $\mathbf{r}_1$, $\mathbf{r}_2$ and $\mathbf{r}_3$ are the position vectors of bodies $m_1$, $m_2$ and $m_3$ respectively, and $\hat{e}_{r_1}$, $\hat{e}_{r_2}$, and $\hat{e}_{r_3}$ are unit vectors along the respective position vectors. Also, $\hat{e}_{\theta_1}$, $\hat{e}_{\theta_2}$ and $\hat{e}_{\theta_3}$ are unit vectors perpendicular to $\hat{e}_{r_1}$, $\hat{e}_{r_2}$, and $\hat{e}_{r_3}$ respectively as clearly indicated by figure 1. The centre of mass of the system is labeled by '$O$'.

Before we present our formal strategy and procedure for solving this problem, it is worth demonstrating, as to why did we specifically develop this method of solution. From figure 1, it is clear that we have attached a frame of reference *oxy* to the centre of mass of the system. We claim that *oxy* is an inertial frame of reference, since it can be shown that the centre of mass of the system has zero acceleration for all time. The derivation of this result is straight forward and we therefore choose to omit its presentation. The use of Newton's 2nd law of motion along with Newton's law for gravitation, allows us to model the system of three bodies as a system of ordinary vector differential equations. This simultaneous system of vector differential equations is presented below

$$\ddot{\mathbf{r}}_1 = \left(\frac{Gm_2}{|\mathbf{r}_2 - \mathbf{r}_1|^2}\right)\left[\frac{\hat{e}_{r_2} - \hat{e}_{r_1}}{|\hat{e}_{r_2} - \hat{e}_{r_1}|}\right] + \left(\frac{Gm_3}{|\mathbf{r}_3 - \mathbf{r}_1|^2}\right)\left[\frac{\hat{e}_{r_3} - \hat{e}_{r_1}}{|\hat{e}_{r_3} - \hat{e}_{r_1}|}\right] \quad (1a)$$

$$\ddot{\mathbf{r}}_2 = -\left(\frac{Gm_1}{|\mathbf{r}_2 - \mathbf{r}_1|^2}\right)\left[\frac{\hat{e}_{r_2} - \hat{e}_{r_1}}{|\hat{e}_{r_2} - \hat{e}_{r_1}|}\right] + \left(\frac{Gm_3}{|\mathbf{r}_3 - \mathbf{r}_2|^2}\right)\left[\frac{\hat{e}_{r_3} - \hat{e}_{r_2}}{|\hat{e}_{r_3} - \hat{e}_{r_2}|}\right] \quad (1b)$$

$$\ddot{\mathbf{r}}_3 = -\left(\frac{Gm_1}{|\mathbf{r}_3 - \mathbf{r}_1|^2}\right)\left[\frac{\hat{e}_{r_3} - \hat{e}_{r_1}}{|\hat{e}_{r_3} - \hat{e}_{r_1}|}\right] - \left(\frac{Gm_2}{|\mathbf{r}_3 - \mathbf{r}_2|^2}\right)\left[\frac{\hat{e}_{r_3} - \hat{e}_{r_2}}{|\hat{e}_{r_3} - \hat{e}_{r_2}|}\right] \quad (1c)$$

The squares of the absolute difference of the position vectors, encountered in (1a), (1b) and (1c) can be expressed as follows, by the use of resolution in the Cartesian coordinate system

$$|\mathbf{r}_2 - \mathbf{r}_1|^2 = r_1^2 + r_2^2 - 2r_1 r_2 \cos(\theta_2 - \theta_1) \quad (2a)$$

$$|\mathbf{r}_3 - \mathbf{r}_1|^2 = r_1^2 + r_3^2 - 2r_1 r_3 \cos(\theta_3 - \theta_1) \quad (2b)$$

$$|\mathbf{r}_3 - \mathbf{r}_2|^2 = r_2^2 + r_3^2 - 2r_2 r_3 \cos(\theta_3 - \theta_2) \quad (2c)$$

We can also derive alternative expressions for the absolute difference of the radial unit vectors, presented as follows

$$|\hat{e}_{r_2} - \hat{e}_{r_1}| = \sqrt{2}\sqrt{1 - \cos(\theta_2 - \theta_1)} \quad (3a)$$

$$|\hat{e}_{r_3} - \hat{e}_{r_1}| = \sqrt{2}\sqrt{1 - \cos(\theta_3 - \theta_1)} \quad (3b)$$

$$|\hat{e}_{r_3} - \hat{e}_{r_2}| = \sqrt{2}\sqrt{1 - \cos(\theta_3 - \theta_2)} \quad (3c)$$

In order to simplify our system of vector differential equations (1a), (1b) and (1c), (or alternatively set (1) ) we can now substitute relation sets (2) and (3) in (1), to get the following simpler system of vector differential equations

$$\ddot{\mathbf{r}}_1 = \left(\frac{Gm_2}{\sqrt{2}[r_1^2 + r_2^2 - 2r_1 r_2 \cos(\theta_2 - \theta_1)]\sqrt{1 - \cos(\theta_2 - \theta_1)}}\right)[\hat{e}_{r_2} - \hat{e}_{r_1}] +$$

$$\left(\frac{Gm_3}{\sqrt{2}[r_1^2 + r_3^2 - 2r_1 r_3 \cos(\theta_3 - \theta_1)]\sqrt{1 - \cos(\theta_3 - \theta_1)}}\right)[\hat{e}_{r_3} - \hat{e}_{r_1}] \quad (4a)$$

$$\ddot{\mathbf{r}}_2 = -\left(\frac{Gm_1}{\sqrt{2}[r_1^2 + r_2^2 - 2r_1 r_2 \cos(\theta_2 - \theta_1)]\sqrt{1 - \cos(\theta_2 - \theta_1)}}\right)[\hat{e}_{r_2} - \hat{e}_{r_1}]$$

$$+ \left(\frac{Gm_3}{\sqrt{2}[r_2^2 + r_3^2 - 2r_2 r_3 \cos(\theta_3 - \theta_2)]\sqrt{1 - \cos(\theta_3 - \theta_2)}}\right)[\hat{e}_{r_3} - \hat{e}_{r_2}] \quad (4b)$$

$$\ddot{\mathbf{r}}_3 = -\left(\frac{Gm_1}{\sqrt{2}[r_1^2+r_3^2-2r_1r_3\cos(\theta_3-\theta_1)]\sqrt{1-\cos(\theta_3-\theta_1)}}\right)\left[\hat{e}_{r_3}-\hat{e}_{r_1}\right]$$

$$-\left(\frac{Gm_2}{\sqrt{2}[r_2^2+r_3^2-2r_2r_3\cos(\theta_3-\theta_2)]\sqrt{1-\cos(\theta_3-\theta_2)}}\right)\left[\hat{e}_{r_3}-\hat{e}_{r_2}\right] \quad (4c)$$

We label this system of equations as system (4), which although seemingly more complicated than (1), is simpler in the sense that it does not involve any absolute terms. This system is therefore easier to handle as compared to (1). Our next step, as should be obvious to the reader, is to resolve each of equations (4a), (4b) and (4c) parallel and perpendicular to unit vectors $\hat{e}_{r_1}$, $\hat{e}_{r_2}$, and $\hat{e}_{r_3}$ respectively, and compare coefficients of the unit vectors on both sides of each equation to derive the associated scalar differential equations. However in order to accomplish this, we must first resolve the unit vectors in terms of each other. Figure 2 emphasizes the arrangement of these unit vectors relative to each other and serves as an aid in their resolution.

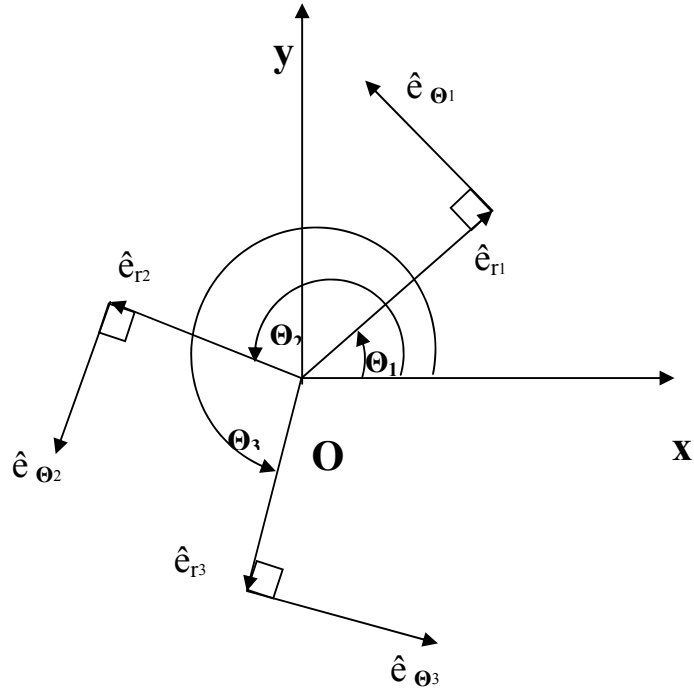

**FIGURE 2**

Making use of figure 2 then, we can resolve two of each of the three radial unit vectors parallel and perpendicular to the remaining single radial unit vector. This results in three systems of relations, each composed of two equations, which are presented as follows. System (5) is obtained by the resolution of $\hat{e}_{r_2}$ and $\hat{e}_{r_3}$ parallel and perpendicular to $\hat{e}_{r_1}$.

$$\hat{e}_{r_2} = \cos(\theta_2 - \theta_1)\hat{e}_{r_1} + \sin(\theta_2 - \theta_1)\hat{e}_{\theta_1} \qquad (5a)$$

$$\hat{e}_{r_3} = \cos(\theta_3 - \theta_1)\hat{e}_{r_1} + \sin(\theta_3 - \theta_1)\hat{e}_{\theta_1} \qquad (5b)$$

Similarly, we can resolve $\hat{e}_{r_1}$ and $\hat{e}_{r_3}$ parallel and perpendicular to $\hat{e}_{r_2}$ to get (6)

$$\hat{e}_{r_1} = \cos(\theta_2 - \theta_1)\hat{e}_{r_2} + \sin(\theta_2 - \theta_1)\hat{e}_{\theta_2} \qquad (6a)$$

$$\hat{e}_{r_3} = \cos(\theta_3 - \theta_2)\hat{e}_{r_2} + \sin(\theta_3 - \theta_2)\hat{e}_{\theta_2} \qquad (6b)$$

As a last step of this procedure, we finally resolve $\hat{e}_{r_1}$ and $\hat{e}_{r_2}$ parallel and perpendicular to $\hat{e}_{r_3}$ to get (7)

$$\hat{e}_{r_1} = \cos(\theta_3 - \theta_1)\hat{e}_{r_3} + \sin(\theta_3 - \theta_1)\hat{e}_{\theta_3} \qquad (7a)$$

$$\hat{e}_{r_2} = \cos(\theta_3 - \theta_2)\hat{e}_{r_3} + \sin(\theta_3 - \theta_2)\hat{e}_{\theta_3} \qquad (7b)$$

As should be obvious by now, we can resolve (4) in polar coordinates with the help of (5), (6) and (7). Doing so and comparing coefficients of the unit vectors on both sides of each equation, we can derive the following associated scalar differential equations as previously mentioned.

$$\ddot{r}_1 - r_1\dot{\theta}_1^{\,2} = \left(\frac{Gm_2[\cos(\theta_2 - \theta_1) - 1]}{\sqrt{2}[r_1^2 + r_2^2 - 2r_1r_2\cos(\theta_2 - \theta_1)]\sqrt{1 - \cos(\theta_2 - \theta_1)}}\right) +$$

$$\left(\frac{Gm_3[\cos(\theta_3 - \theta_1) - 1]}{\sqrt{2}[r_1^2 + r_3^2 - 2r_1r_3\cos(\theta_3 - \theta_1)]\sqrt{1 - \cos(\theta_3 - \theta_1)}}\right) \qquad (8a)$$

$$\ddot{r}_1 - r_1\dot{\theta}_1^{\,2} = \left(\frac{Gm_2[\cos(\theta_2 - \theta_1) - 1]}{\sqrt{2}[r_1^2 + r_2^2 - 2r_1r_2\cos(\theta_2 - \theta_1)]\sqrt{1 - \cos(\theta_2 - \theta_1)}}\right) +$$

$$\left(\frac{Gm_3[\cos(\theta_3 - \theta_1) - 1]}{\sqrt{2}[r_1^2 + r_3^2 - 2r_1r_3\cos(\theta_3 - \theta_1)]\sqrt{1 - \cos(\theta_3 - \theta_1)}}\right) \qquad (8b)$$

$$\ddot{r}_2 - r_2\dot{\theta}_2^{\,2} = \left(\frac{Gm_1[\cos(\theta_2 - \theta_1) - 1]}{\sqrt{2}[r_1^2 + r_2^2 - 2r_1r_2\cos(\theta_2 - \theta_1)]\sqrt{1 - \cos(\theta_2 - \theta_1)}}\right) +$$

$$\left(\frac{Gm_3[\cos(\theta_3 - \theta_2) - 1]}{\sqrt{2}[r_2^2 + r_3^2 - 2r_2r_3\cos(\theta_3 - \theta_2)]\sqrt{1 - \cos(\theta_3 - \theta_2)}}\right) \qquad (8c)$$

$$r_2\ddot{\theta}_2 + 2\dot{r}_2\dot{\theta}_2 = -\left(\frac{Gm_1\sin(\theta_2 - \theta_1)}{\sqrt{2}[r_1^2 + r_2^2 - 2r_1r_2\cos(\theta_2 - \theta_1)]\sqrt{1 - \cos(\theta_2 - \theta_1)}}\right)$$

$$-\left(\frac{Gm_3 \sin(\theta_3 - \theta_2)}{\sqrt{2}[r_2^2 + r_3^2 - 2r_2r_3\cos(\theta_3 - \theta_2)]\sqrt{1 - \cos(\theta_3 - \theta_2)}}\right) \quad (8d)$$

$$\ddot{r_3} - r_3\dot{\theta_3}^2 = \left(\frac{Gm_1[\cos(\theta_3 - \theta_1) - 1]}{\sqrt{2}[r_1^2 + r_3^2 - 2r_1r_3\cos(\theta_3 - \theta_1)]\sqrt{1 - \cos(\theta_3 - \theta_1)}}\right) +$$

$$\left(\frac{Gm_2[\cos(\theta_3 - \theta_2) - 1]}{\sqrt{2}[r_2^2 + r_3^2 - 2r_2r_3\cos(\theta_3 - \theta_2)]\sqrt{1 - \cos(\theta_3 - \theta_2)}}\right) \quad (8e)$$

$$r_3\ddot{\theta_3} + 2\dot{r_3}\dot{\theta_3} = -\left(\frac{Gm_1 \sin(\theta_3 - \theta_1)}{\sqrt{2}[r_1^2 + r_3^2 - 2r_1r_3\cos(\theta_3 - \theta_1)]\sqrt{1 - \cos(\theta_3 - \theta_1)}}\right)$$

$$-\left(\frac{Gm_2 \sin(\theta_3 - \theta_2)}{\sqrt{2}[r_2^2 + r_3^2 - 2r_2r_3\cos(\theta_3 - \theta_2)]\sqrt{1 - \cos(\theta_3 - \theta_2)}}\right) \quad (8f)$$

It should be noted here that system (8) is the scalar analogue of our original system (1). We believe that the mere sight of (8) is sufficient to demonstrate the magnitude of the difficulties that would be encountered in solving this problem, since a solution of the three body problem would require that we solve system (8), not by numerical techniques, but by exact mathematical treatments. We therefore reject the idea of adopting such a straightforward strategy in solving this problem. It is important to realize at this stage that such a problem might be simpler to solve (on a relative scale, that is) by using some other strategy that is not as direct as the one demonstrated above. Moreover, we might as well drop the idea of looking for exact solutions, since closed form approximations with a reasonable accuracy might just be useful. Having adequately provided the motivation behind our method of solution, we go on to present our formal procedure for finding closed form approximation solutions for the Three Body Problem.

**Main Results**

It should be obvious by now that we need more simplifications for finding a solution. Therefore, we make another assumption in addition to the ones proposed in the introduction. A brief discussion on this assumption is what follows.

We specifically assume that while executing three body motion, each body remains approximately in two body motion with a body having the sum of the masses of the other two bodies, placed at the centre of mass of these two bodies. Stated in a more comprehensive manner, we propose the following three statements, which of course, are explanatory versions of our assumption:

(1)    $m_1$ approximately remains in two body motion with a body of mass $m_2 + m_3$ placed at the centre of mass of $m_2$ and $m_3$ given by

$$\mathbf{r}_{23} = \frac{m_2 \mathbf{r}_2 + m_3 \mathbf{r}_3}{m_2 + m_3} \quad \forall \ t$$

(2)    $m_2$ approximately remains in two body motion with a body of mass $m_1 + m_3$ placed at the centre of mass of $m_1$ and $m_3$ given by

$$\mathbf{r}_{13} = \frac{m_1 \mathbf{r}_1 + m_3 \mathbf{r}_3}{m_1 + m_3} \quad \forall \ t$$

(3)    $m_3$ approximately remains in two body motion with a body of mass $m_1 + m_2$ placed at the centre of mass of $m_1$ and $m_2$ given by

$$\mathbf{r}_{12} = \frac{m_1 \mathbf{r}_1 + m_2 \mathbf{r}_2}{m_1 + m_2} \quad \forall \ t$$

The key point here is to note how we attempt to replicate the effect of two individual bodies on each body. Say as a further explanation for statement 1 above, we try replicating the individual gravitational effects of $m_2$ *and* $m_3$ on $m_1$, simply by placing a body mass $m_2 + m_3$ at the centre of mass of $m_2$ and $m_3$. We claim that the motion of $m_1$ remains more or less the same, be it two body motion between $m_1$ and $m_2 + m_3$, or three body motion between $m_1$, $m_2$ *and* $m_3$. We will use the two body motion analogue just explained, to make predictions relating to three body motion. This of course could be an accountable source of error in the solutions that we will compute. The proposed assumption, though slightly inaccurate, cannot be regarded as being invalid, since our aim in the first place was to find closed form approximations, and not exact solutions. Moreover, had this assumption been exactly valid, the three body problem would have been reducible to the two body problem, and finding exact solutions would not have been that difficult. We no longer comment on the validity of our new assumption, and go on with its application in computing the solutions. The following three figures, figures 3, 4 and 5 provide a diagrammatic illustration of the applications of statements 1, 2 and 3 respectively, to our original problem of three bodies.

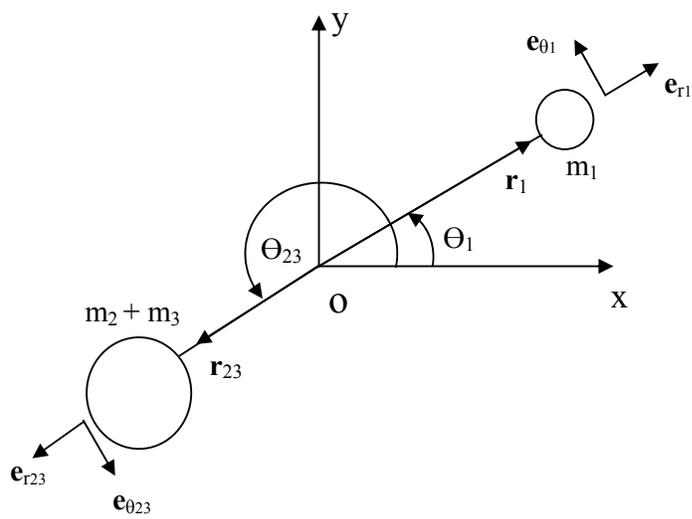

**Figure 3**

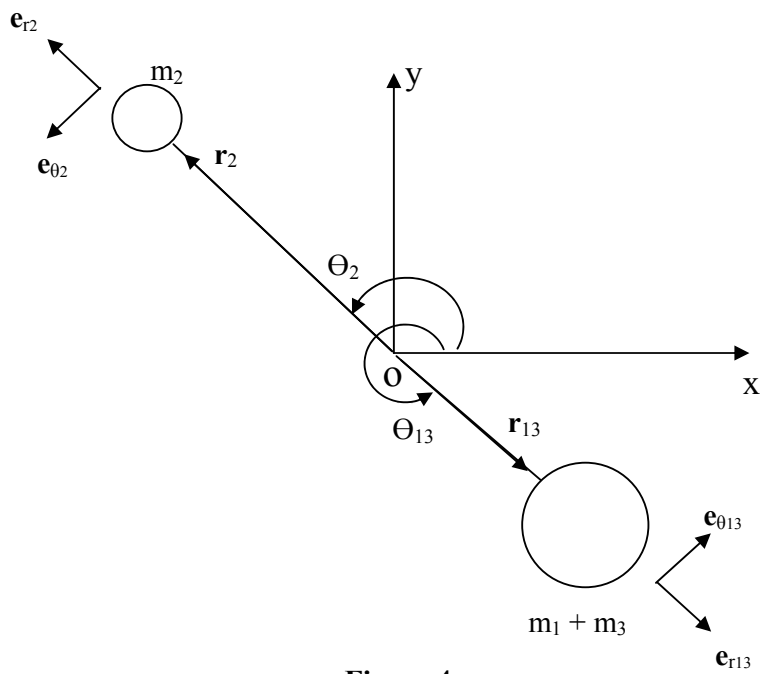

**Figure 4**

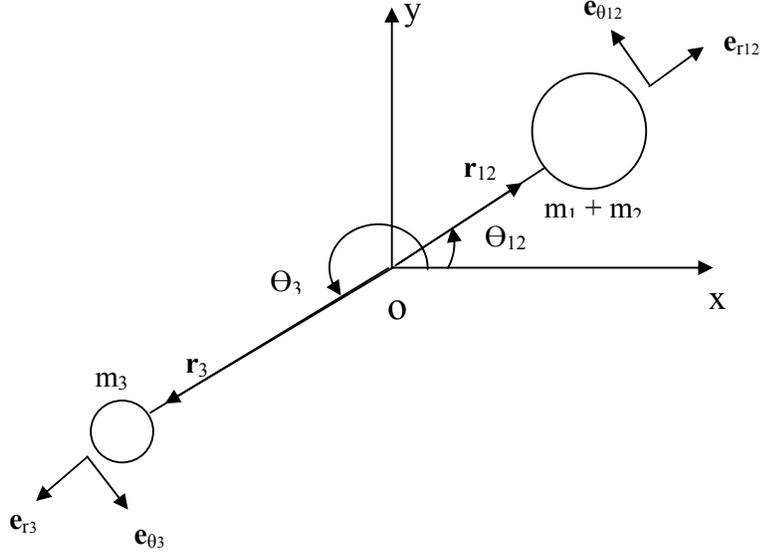

**Figure 5**

We will make use of the configurations shown in figures 3, 4 and 5 to solve for the position vectors $\mathbf{r}_1$, $\mathbf{r}_2$ and $\mathbf{r}_3$ as functions of time respectively. The additional unit vectors required for $\mathbf{r}_{23}$, $\mathbf{r}_{13}$ and $\mathbf{r}_{12}$ are shown in figures 3, 4 and 5 respectively. Also, we choose to represent $\mathbf{r}_{23}$, $\mathbf{r}_{13}$ and $\mathbf{r}_{12}$ as $r_{23}(t)\hat{e}_{r_{23}}$, $r_{13}(t)\hat{e}_{r_{13}}$ and $r_{12}(t)\hat{e}_{r_{12}}$ respectively. This matches with the notation that we used previously for $\mathbf{r}_1$, $\mathbf{r}_2$ and $\mathbf{r}_3$. We now explain some crucial ideas from figure 3, which can be readily extended to figures 4 and 5.

From figure 3 then, in accordance with statement (1), unit vectors $\hat{e}_{r_1}$ and $\hat{e}_{r_{23}}$ remain approximately collinear for all time i.e. $\hat{e}_{r_1} \cdot \hat{e}_{r_{23}} \approx -1 \; \forall \; t$. Also, we define the vector $\mathbf{r}_a = \mathbf{r}_1 - \mathbf{r}_{23}$, from which it follows that the following relations hold true

$$\mathbf{r}_a = \mathbf{r}_1 - \mathbf{r}_{23} \tag{9a}$$

$$\hat{e}_{r_a} = \hat{e}_{r_1} = -\hat{e}_{r_{23}} \tag{9b}$$

$$\hat{e}_{\theta_a} = \hat{e}_{\theta_1} = -\hat{e}_{\theta_{23}} \tag{9c}$$

$$r_a(t) = r_1(t) + r_{23}(t) \tag{9d}$$

$$\theta_a(t) = \theta_1(t) = \theta_{23}(t) + \pi \tag{9e}$$

$$\dot{\theta}_a(t) = \dot{\theta}_1(t) = \dot{\theta}_{23}(t) \tag{9f}$$

Note that in $(9d)$, whether we add or subtract $\pi$, it is essentially the same thing.

Modeling the system in figure 3 then, by the use of Newton's second law and Newton's law for gravitation, and making use of $(9b)$ and $(9d)$, we get

$$\ddot{\mathbf{r}}_1 = -\left(\frac{G(m_2 + m_3)}{r_a^2}\right)\hat{e}_{r_a} \tag{10a}$$

$$\ddot{\mathbf{r}}_{23} = \left(\frac{Gm_1}{r_a^2}\right)\hat{e}_{r_a} \tag{10b}$$

We could now make use of $(9a)$ to express this model in a more compact form

$$\ddot{\mathbf{r}}_a = -\left(\frac{G(m_1 + m_2 + m_3)}{r_a^2}\right)\hat{e}_{r_a} \tag{10c}$$

Resolving $(10c)$ in polar coordinates and comparing coefficients of the unit vectors on both sides of the equation, we can obtain the following scalar analogue of $(10c)$

$$\ddot{r}_a - r_a \dot{\theta}_a^{\,2} = -\left(\frac{G(m_1 + m_2 + m_3)}{r_a^2}\right) \tag{11a}$$

$$r_a \ddot{\theta}_a + 2\dot{r}_a \dot{\theta}_a = 0 \tag{11b}$$

We could use system (11) to exactly solve for $r_a(\theta_a)$. However our requirement is different and this is the point where our assumptions come into play. We argued at the beginning that the angular velocities involved were smaller than 1 radians per second, and that the position vectors were not arbitrarily large. In the following we provide a mathematical representation of this fact.

$$|\dot{\theta}_1| < 1 \; rad/s \tag{12a}$$

$$|\dot{\theta}_2| < 1 \; rad/s \tag{12b}$$

$$|\dot{\theta}_3| < 1 \; rad/s \tag{12c}$$

$$|\mathbf{r}_1| << \infty; \text{ or equivalently}: r_1(t) << \infty \tag{12d}$$

$$|\mathbf{r}_2| << \infty; \text{ or equivalently}: r_2(t) << \infty \tag{12e}$$

$$|\mathbf{r}_3| << \infty; \text{ or equivalently}: r_3(t) << \infty \tag{12f}$$

Considering the expression defining $\mathbf{r}_{23}$ in statement number 1 mentioned at the beginning of this section, and taking into account (12e) and (12f), we can easily conclude that $r_{23}(t) << \infty$. Taking this fact and $(9d)$ into account, it is more than a simple task to conclude that $r_a(t) << \infty$. Use of this information along with $(9e)$ allows us to set $r_a \dot{\theta}_a^2 \simeq 0$ in equation (11a), which then takes the form

$$\ddot{r}_a = -\left(\frac{G(m_1 + m_2 + m_3)}{r_a^2}\right) \quad (13a)$$

Also, multiplying (11b) by '$r_a$' and recognizing the left hand side of the resulting equation as the expansion of the product rule for derivatives, we get $d(r_a^2 \dot{\theta}_a) = 0$, integration of which yields

$$r_a^2 \dot{\theta}_a = r_{ao}^2 \dot{\theta}_{ao} \quad (13b)$$

Here $r_{ao} = r_{1o} = r_1(0)$, and $\dot{\theta}_{ao} = \dot{\theta}_{1o} = \dot{\theta}_1(0)$. As a next step, multiplication of (13a) by '$\dot{r}_a \, dt$' allows us to integrate the resulting equation on the left hand side w.r.t. '$t$' and on the right hand side w.r.t. '$r$', and simplification yields the form

$$\frac{dr_a}{dt} = \pm\left[\frac{A_a}{r} + B_a\right]^{\frac{1}{2}} \quad (13c)$$

where $A_a = 2G(m_1 + m_2 + m_3)$ and $B_a = \dot{r}_{ao}^2 - \frac{2G(m_1 + m_2 + m_3)}{r_{ao}}$. As a next step, we seperate variables and apply integration to both sides of the equation to get

$$\int_{r_{ao}}^{r_a}\left[\frac{A_a}{B_a r_a} + 1\right]^{-\frac{1}{2}} dr_a = \pm\sqrt{B_a}\int_{t_o}^{t} dt \quad (13d)$$

Note that $B_a \geq 0$ in the above relation. At the end, we will sum up the conditions for which our solutions hold true. Now, although (13d) can be integrated in its current form and an implicit equation relating $r_a$ and $t$ can be found, however $r_a(t)$ cannot be explicitly solved for. This fact encourages us to try a simple binomial approximation of the form $\left[\frac{A_a}{B_a r_a} + 1\right]^{-\frac{1}{2}} \simeq 1 - \left(\frac{A_a}{2 B_a r_a}\right) \quad \forall \quad |r_a| > \left|\frac{A_a}{B_a}\right|$. Use of this approximation simplifies (13d) so that we are capable of integrating on both sides and deriving the following implicit equation

$$r_a + \ln r_a^{-k_a} = f(t) \quad (13e)$$

where $k_a = \left(\frac{A_a}{2B_a}\right)$ and $f(t) = \pm\sqrt{B_a}(t - t_o) + r_{ao} - \ln r_{ao}^{k_a}$. Here again, it should be noted that $B_a$ should not be allowed to attain negative values for the validity of result (13e). As a reminder, we mention here again that we will later on collect all the requirements that should hold for our solutions to be valid. Therefore the reader should not bother at this stage about the solvability conditions of the problem. Now, solving for $r_a(t)$ explicitly from (13e) we get

$$r_a(t) = -k_a * lambertw\left[\frac{-e^{-\frac{f}{k_a}}}{k_a}\right] \quad (14)$$

where '$lambertw$' is the notation used for the lambert's wave function. Replacing the expressions for $k_a$ and $f(t)$ into (14) and performing a few straight forward manipulations, we can derive the expression

$$r_a(t) = -\left(\frac{A_a}{2B_a}\right) lambertw\left[c_{4a} e^{c_{5a} t}\right] \quad (15)$$

where $c_{4a} = c_{1a} e^{c_{2a} t_o}$, $c_{5a} = -c_{2a}$, $c_{1a} = -\left(\frac{2B_a}{A_a}\right) e^{(-\frac{2B_a}{A_a})(r_{ao} - \ln r_{ao}^{\frac{A_a}{2B_a}})}$ and $c_{2a} = \pm\left(\frac{2B_a \sqrt{B_a}}{A_a}\right)$.

Having computed $r_a(t)$, we will now make use of relation (10a) to find $r_1(t)$ and $\theta_1(t)$. Also, in a manner similar to the one adopted in the derivation of (13a) and (13b) from (10c), we can derive the following set of relations from (10c).

$$\ddot{r_1} = -\left(\frac{G(m_2 + m_3)}{r_a^2(t)}\right) \quad (16a)$$

$$r_1^2 \dot{\theta_1} = r_{1o}^2 \dot{\theta_{1o}} \quad (16b)$$

We now substitute (15) in (16a) and integrate twice w.r.t. $t$ to get our first closed form approximation for $r_1(t)$.

$$r_1(t) = k_{aa} t + \left(\frac{k_{ba}}{c_{5a}}\right)\left[\frac{1}{2}(lambertw[c_{4a} e^{c_{5a} t}])^4 + (lambertw[c_{4a} e^{c_{5a} t}])^3 + \frac{1}{2}(lambertw[c_{4a} e^{c_{5a} t}])^2\right]$$

$$- k_{aa} t_o - \left(\frac{k_{ba}}{c_{5a}}\right)\left[\begin{array}{c} \frac{1}{2}(lambertw[c_{4a} e^{c_{5a} t_o}])^4 + \\ (lambertw[c_{4a} e^{c_{5a} t_o}])^3 + \frac{1}{2}(lambertw[c_{4a} e^{c_{5a} t_o}])^2 \end{array}\right] + r_{1o} \quad (17a)$$

where $r_{1o} = r_1(0)$, $\dot{r_{1o}} = \dot{r_1}(0)$, $k_{aa} = \dot{r_{1o}} - \left(\frac{k_{1a}}{2c_{5a}}\right)\left[1 + 2lambertw(c_{4a} e^{c_{5a} t_o})\right]$,

$$k_{ba} = \left(\frac{k_{1a}}{2c_{5a}}\right) \text{ and } k_{1a} = -\left(\frac{4B_a^2 G(m_2 + m_3)}{A_a^2}\right)$$ (here all the right hand side constants have been previously defined)

Having accomplished the above result, we are now free to substitute equation (17a) in (16b), separate variables, and integrate w.r.t. $t$ to obtain the following explicit solution for $\theta_1(t)$

$$\theta_1(t) = \theta_{1o} + \left(\frac{4r_{ao}^2 \dot{\theta}_{1o} B_a^2}{c_{5a} A_a^2}\right)(1 + 2\,lambertw[c_{4a} e^{c_{5a} t_o}])(lambertw[c_{4a} e^{c_{5a} t_o}])^2$$

$$- \left(\frac{4r_{ao}^2 \dot{\theta}_{1o} B_a^2}{c_{5a} A_a^2}\right)(1 + 2\,lambertw[c_{4a} e^{c_{5a} t}])(lambertw[c_{4a} e^{c_{5a} t}])^2 \quad (17b)$$

Most of our work is essentially complete since we've successfully approximated the motion of body $m_1$, given by (17a) and (17b). As mentioned previously, we go on to sum up the conditions for which this solution is valid. First of all, we require that relation set (12) holds true. Secondly, it should be quite obvious to the reader that this solution is rendered infeasible in case of collision or explosion analysis.

Recall that the condition required for the validity of the binomial approximation used to simplify (13d) was $|\mathbf{r}_a| > \left|\frac{A_a}{B_a}\right|$. Also, a careful look at the various equations encountered while solving the problem, should help us to easily conclude that we also require $B_a$ to be positive. This task, being trivially simple, has been consciously left out for the reader to figure out. Therefore, we require $B_a > 0$. Also, from the definitions of $\mathbf{r}_a$, $A_a$ and $B_a$ then, we can translate the conditions mentioned in this paragraph into complicated expressions, simplification of which would require a rather long discussion. As a hint we state that in order to translate $|\mathbf{r}_a| > \left|\frac{A_a}{B_a}\right|$ in terms of our elementary variables $r_1$, $r_2$, $r_3$, $\theta_1$, $\theta_2$, and $\theta_3$, we need to realize this expression as $\left|r_1(t)\hat{e}_{r_1} - \left(\frac{m_2 r_2(t)\hat{e}_{r_2} + m_3 r_3(t)\hat{e}_{r_3}}{m_2 + m_3}\right)\right| > \left|\frac{A_a}{B_a}\right|$ and then use the definitions of $A_a$ and $B_a$ and the resolution of the unit vectors in terms of each other.

However we choose to omit this task and keep things as simple as possible. Summing up, we state that in order for (17a) and (17b) to hold true, we require that the following condition holds true in addition to relation set (12)

$$|\mathbf{r}_a| > \left|\frac{A_a}{B_a}\right| \text{ and } B_a > 0 \tag{18}$$

Note that by computing (17a) and (17b) we have infact analytically solved for $\mathbf{r}_1(t)$. The task of finding $\mathbf{r}_2(t)$ and $\mathbf{r}_3(t)$ is trivial in the sense that it requires us to follow a sequence of steps, similar to the ones presented for computing (17a) and (17b), and hence we will not hesitate in directly presenting the solutions. The configuration needed for finding $r_2(t)$ and $\theta_2(t)$ is shown in figure 4. Proceeding in exactly a similar fashion as we did before, we can show that

$$r_2(t) = k_{ab}t + \left(\frac{k_{bb}}{c_{5b}}\right)\left[\frac{1}{2}(lambertw[c_{4b}e^{c_{5b}t}])^4 + (lambertw[c_{4b}e^{c_{5b}t}])^3 + \frac{1}{2}(lambertw[c_{4b}e^{c_{5b}t}])^2\right]$$

$$-k_{ab}t_o - \left(\frac{k_{bb}}{c_{5b}}\right)\left[\begin{array}{c}\frac{1}{2}(lambertw[c_{4b}e^{c_{5b}t_o}])^4 + \\ (lambertw[c_{4b}e^{c_{5b}t_o}])^3 + \frac{1}{2}(lambertw[c_{4b}e^{c_{5b}t_o}])^2\end{array}\right] + r_{2o} \tag{19a}$$

$$\theta_2(t) = \theta_{2o} + \left(\frac{4r_{bo}^2 \dot{\theta}_{2o} B_b^2}{c_{5b}A_b^2}\right)(1 + 2lambertw[c_{4b}e^{c_{5b}t_o}])(lambertw[c_{4b}e^{c_{5b}t_o}])^2$$

$$- \left(\frac{4r_{bo}^2 \dot{\theta}_{2o} B_b^2}{c_{5b}A_b^2}\right)(1 + 2lambertw[c_{4b}e^{c_{5b}t}])(lambertw[c_{4b}e^{c_{5b}t}])^2 \tag{19b}$$

where $r_{2o} = r_2(0)$, $\dot{r}_{2o} = \dot{r}_2(0)$, $k_{ab} = \dot{r}_{2o} - \left(\frac{k_{1b}}{2c_{5b}}\right)\left[1 + 2lambertw(c_{4b}e^{c_{5b}t_o})\right]$,

$k_{bb} = \left(\frac{k_{1b}}{2c_{5b}}\right)$, $k_{1b} = -\left(\frac{4B_b^2 G(m_1 + m_3)}{A_b^2}\right)$, $c_{4b} = c_{1b}e^{c_{2b}t_o}$, $c_{5b} = -c_{2b}$,

$c_{1b} = -\left(\frac{2B_b}{A_b}\right)e^{(-\frac{2B_b}{A_b})(r_{bo} - \ln r_{bo}^{\frac{A_b}{2B_b}})}$, $c_{2b} = \pm\left(\frac{2B_b\sqrt{B_b}}{A_b}\right)$, $A_b = 2G(m_1 + m_2 + m_3)$ and lastly,

$B_b = \dot{r}_{bo}^2 - \left(\frac{2G(m_1 + m_2 + m_3)}{r_{bo}}\right)$. Note that (19a) and (19b) are valid, provided that the following condition holds in addition to relation set (12)

$$|\mathbf{r}_b| > \left|\frac{A_b}{B_b}\right| \text{ and } B_b > 0 \tag{20}$$

Having presented our results for $r_2(t)$ and $\theta_2(t)$, we merely state once again, that $r_3(t)$ and $\theta_3(t)$ (presented in what follows) can be derived by following a similar line of mathematical arguments and steps. The configuration that needs to be considered for this task is shown in figure 5. In the following, we directly present our results.

$$r_3(t) = k_{ac}t + \left(\frac{k_{bc}}{c_{5c}}\right)\left[\frac{1}{2}(lambertw[c_{4c}e^{c_{5c}t}])^4 + (lambertw[c_{4c}e^{c_{5c}t}])^3 + \frac{1}{2}(lambertw[c_{4c}e^{c_{5c}t}])^2\right]$$

$$-k_{ac}t_o - \left(\frac{k_{bc}}{c_{5c}}\right)\left[\begin{array}{l}\frac{1}{2}(lambertw[c_{4c}e^{c_{5c}t_o}])^4 + \\ (lambertw[c_{4c}e^{c_{5c}t_o}])^3 + \frac{1}{2}(lambertw[c_{4c}e^{c_{5c}t_o}])^2\end{array}\right] + r_{3o} \tag{21a}$$

$$\theta_3(t) = \theta_{3o} + \left(\frac{4r_{co}^2 \dot{\theta}_{3o} B_c^2}{c_{5c}A_c^2}\right)(1 + 2lambertw[c_{4c}e^{c_{5c}t_o}])(lambertw[c_{4c}e^{c_{5c}t_o}])^2$$

$$- \left(\frac{4r_{co}^2 \dot{\theta}_{3o} B_c^2}{c_{5c}A_c^2}\right)(1 + 2lambertw[c_{4c}e^{c_{5c}t}])(lambertw[c_{4c}e^{c_{5c}t}])^2 \tag{21b}$$

where $r_{3o} = r_3(0)$, $\dot{r}_{3o} = \dot{r}_3(0)$, $k_{ac} = \dot{r}_{3o} - \left(\frac{k_{1c}}{2c_{5c}}\right)\left[1 + 2lambertw(c_{4c}e^{c_{5c}t_o})\right]$,

$k_{bc} = \left(\frac{k_{1c}}{2c_{5c}}\right)$, $k_{1c} = -\left(\frac{4B_c^2 G(m_1 + m_2)}{A_c^2}\right)$, $c_{4c} = c_{1c}e^{c_{2c}t_o}$, $c_{5c} = -c_{2c}$,

$c_{1c} = -\left(\frac{2B_c}{A_c}\right)e^{(-\frac{2B_c}{A_c})(r_{co} - \ln r_{co}^{\frac{A_c}{2B_c}})}$, $c_{2c} = \pm\left(\frac{2B_c\sqrt{B_c}}{A_c}\right)$, $A_c = 2G(m_1 + m_2 + m_3)$ and lastly,

$B_c = \dot{r}_{co}^2 - \left(\frac{2G(m_1 + m_2 + m_3)}{r_{co}}\right)$. Note that (21a) and (21b) are valid, provided that the following condition holds in addition to relation set (12)

$$|\mathbf{r}_c| > \left|\frac{A_c}{B_c}\right| \text{ and } B_c > 0 \tag{22}$$

We now sum up our discussion, stating that by finding $r_2(t)$ and $\theta_2(t)$, we infact completely described the vector $\mathbf{r}_2$ as a time function. This in turn implies that we were successful in approximating the motion of $m_2$. Similarly, the derivation of $r_3(t)$ and $\theta_3(t)$ enables us to describe the motion of $m_3$ explicitly as a function of time. The discussion on our methodology for solving the problem is essentially complete, and we now go on to present the summary.

**Summary:**

Having presented a formal discussion on the solution procedure developed in this paper, we believe that the solutions are accurate enough for predicting the future positions of the three masses, provided that validity conditions for our solutions hold true. As should be clear by now, the approach has been simple but required extensive mathematical manipulations. Using Newton's laws for gravitation and resultant acceleration, we modeled a two body motion analogue system for our problem of three bodies (this will later be discussed). We then used the assumption that the angular velocities involved were considerably less than one radians per second, and that the position vectors involved were not arbitrarily large. The assumptions proposed were both simple and practically feasible, apart from one, which we will shortly discuss. It was through the use of these assumptions that we were able to simplify our problem and obtain solutions, given by (17a), (17b), (19a), (19b), (21a) and (21b). We also developed the solvability conditions for our solutions, which were given by relation set (12), (18), (20) and (22). One of our assumptions that we believe might cause apprehension, has been discussed in what follows. We assumed that while performing three body motion, each body specifically remains in two body motion with a body having the sum of the masses of the other two bodies, placed at the centre of these two bodies. The reader should refer back to statements (1), (2) and (3) at the beginning of the main results. We identify this assumption as the main source of error in the solutions that we've formulated. The reason behind this, being trivially simple, is that three body motion is chaotic, whereas two body motion is non chaotic. Therefore, a two body motion replication of three body motion is inexact. As an explanation for taking this assumption, we claim here once again, that our aim in the first place was to develop a procedure for finding closed form approximations. Doubtlessly, the Poincare's Dictum comprehensively proves that the problem is truly insolvable as a result of the nature of the instabilities involved.

# References

[1]  R. Broucke and D. Boggs, Periodic orbits in the planar general 3-body problem, Celestial Mechanics, Vol. II, (1975) 13-38.

[2]  P. J. Shelus, A two parameter survey of periodic orbits in the restricted problem of 3 bodies, Celestial Mechanics, Vol. 5 (1972) 483-489.

[3]  V. V. Markellos, Numerical investigation of the planar restricted three-body problem, Celestial Mechanics, Vol. 9 (1974) 365-380.

[4]  L. M. Perko, Periodic orbits in the restricted 3-body problem: existence and asymptotic approximation, Siam Journal of Applied Mathematics, Vol. 27 (1974) 200.

[5]  M. Bruschi and F. Calogero, Solvable and/or integrable and/or linearizable N-body problems in ordinary (three dimensional) space, J. Nonlinear Math. Phys. Vol. 7, (2000) 303-306.

[6]  F. Calogero, A solvable many body problem in the plane, J. Nonlinear Math. Phys. Vol. 5, (1998) 289-293.